\documentclass[
aps,
prl,
tightenlines,
superscriptaddress,
nofootinbib,
floatfix,
showpacs,
groupedaddress,
preprintnumbers,
nofootinbib,
twocolumn,
amsmath,
amsfonts,
amssymb]
{revtex4}
\usepackage{graphicx,
longtable,
color
}

\begin{document}

\preprint{IP/BBSR/2016-6}

\title{Octant of $\theta_{23}$ in danger with a light sterile neutrino}%

\author{		Sanjib Kumar Agarwalla}\email{sanjib@iopb.res.in}
\affiliation{		Institute of Physics, Sachivalaya Marg, Sainik School Post, Bhubaneswar 751005, India}
\affiliation{		Homi Bhabha National Institute, Training School Complex, Anushakti Nagar, Mumbai 400085, India}

\author{		Sabya Sachi Chatterjee}\email{sabya@iopb.res.in}
\affiliation{		Institute of Physics, Sachivalaya Marg, Sainik School Post, Bhubaneswar 751005, India}
\affiliation{		Homi Bhabha National Institute, Training School Complex, Anushakti Nagar, Mumbai 400085, India}

\author{		Antonio Palazzo}\email{palazzo@ba.infn.it}
\affiliation{ 	Dipartimento Interateneo di Fisica ``Michelangelo Merlin,'' Via Amendola 173, 70126 Bari, Italy}
\affiliation{ 	Istituto Nazionale di Fisica Nucleare, Sezione di Bari, Via Orabona 4, 70126 Bari, Italy}


\begin{abstract}
Present global fits of world neutrino data hint towards non-maximal $\theta_{23}$ with two nearly degenerate
solutions, one in the lower octant ($\theta_{23} <\pi/4$), and the other in the higher octant ($\theta_{23} >\pi/4$).
This octant ambiguity of $\theta_{23}$ is one of the fundamental issues in the neutrino sector, and its
resolution is a crucial goal of next-generation long-baseline (LBL) experiments. In this letter, we address for the first time, 
the impact of a light eV-scale sterile neutrino towards such a measurement, taking the Deep Underground  
Neutrino Experiment (DUNE) as a case study. In the so-called 3+1 scheme involving three active and one sterile
neutrinos, the $\nu_\mu \to \nu_e$ transition probability probed in the LBL experiments acquires a new interference 
term via active-sterile oscillations. We find that this interference term can mimic a swap of the $\theta_{23}$ octant, 
even if one uses the information from both neutrino and antineutrino channels. As a consequence, 
the sensitivity to the octant of $\theta_{23}$ can be completely lost, and this may have serious implications 
for our understanding of neutrinos from both the experimental and theoretical perspectives.
 
\end{abstract}
\maketitle

{\bf {\em Introduction.}}  After the discovery of the smallest leptonic mixing angle $\theta_{13}$,
the 3-flavor paradigm has been firmly established, and neutrino physics is entering the precision era.
In the standard 3-flavor framework, the oscillations are governed by two  mass-squared splittings
$\Delta m^2_{21} \equiv m_2^2-m_1^2$, and $\Delta m^2_{31} \equiv m_3^2- m_1^2$, 
and three mixing angles ($\theta_{12}, \theta_{23}, \theta_{13}$).
Two fundamental elements are still missing in this picture: the mass hierarchy%
\footnote{Here, MH refers to the sign of $\Delta m^2_{31}$, which can be positive
termed as normal hierarchy (NH), or negative denoted as inverted hierarchy (IH).}
(MH) and the CP-phase $\delta$, which, if  $\ne (0,\pi)$,  gives rise to leptonic CP-violation (CPV). 
A rich experimental program is underway to identify these two
unknown properties, and to refine the estimates of the known mass-mixing parameters~\cite{Pascoli:2013wca,Agarwalla:2013hma,Agarwalla:2014fva,Feldman:2012qt,Stanco:2015ejj}.
 
Current global neutrino data hint towards 
non-maximal $\theta_{23}$ with two nearly degenerate solutions: one $< \pi/4$,
termed as lower octant (LO), and the other $>\pi/4$, denoted as higher octant (HO).
The identification of the $\theta_{23}$ octant is a crucial goal in neutrino physics 
due to its deep implications for the theory of neutrino masses and mixing 
(see~\cite{Mohapatra:2006gs,Albright:2006cw,Altarelli:2010gt,King:2014nza,King:2015aea} for reviews).  
Notable models where the $\theta_{23}$ octant plays a key role are 
$\mu \leftrightarrow \tau$ symmetry\footnote{As shown in~\cite{Merle:2014eja,Rivera-Agudelo:2015vza}, 
in the 3+1 scheme, the condition $\theta_{23} \simeq 45^{\circ}$ implies an approximate realization 
of $\mu \leftrightarrow \tau$ symmetry, similarly to what happens in the standard 3-flavor scheme. 
Therefore, establishing that $\theta_{23}$ is maximal (non-maximal) would imply that 
$\mu \leftrightarrow \tau$ symmetry is unbroken (broken), irrespective of the existence 
of a light sterile neutrino.}~\cite{Fukuyama:1997ky,Mohapatra:1998ka,Lam:2001fb,Harrison:2002et,Kitabayashi:2002jd,Grimus:2003kq,Koide:2003rx,Mohapatra:2005yu}, $A_4$ flavor symmetry \cite{Ma:2002ge,Ma:2001dn,Babu:2002dz,Grimus:2005mu,Ma:2005mw}, quark-lepton complementarity~\cite{Raidal:2004iw,Minakata:2004xt,Ferrandis:2004vp,Antusch:2005ca}, and neutrino mixing anarchy~\cite{Hall:1999sn,deGouvea:2012ac}. From a phenomenological perspective, the information on the $\theta_{23}$ octant is also a vital input. In fact, it is well known that the identification of the two unknown properties (MH and CPV) is strictly  intertwined with the determination of $\theta_{23}$ because of parameter degeneracy issues~\cite{Fogli:1996pv,Barger:2001yr,Minakata:2001qm,Hiraide:2006vh,BurguetCastell:2002qx,Agarwalla:2013ju,Machado:2013kya,Minakata:2013hgk,Chatterjee:2013qus}.  

One of the most promising options to identify the $\theta_{23}$ octant is offered by the  
long-baseline (LBL) accelerator experiments. In these setups, a synergy between the 
$\nu_\mu \to \nu_e$ appearance and $\nu_\mu \to \nu_\mu$ disappearance 
channels exists~\cite{Fogli:1996pv,Hiraide:2006vh}, which confers an enhanced sensitivity 
to the $\theta_{23}$ octant. The  $\nu_\mu \to \nu_\mu$ survival probability, at the leading order, 
depends on $\sin^2 2\theta_{23}$. Then, it is sensitive to deviations from maximality but is insensitive 
to the octant. On the other hand, the leading contribution to the $\nu_\mu \to \nu_e$ probability 
is proportional to $\sin^2 \theta_{23}$, and is thus sensitive to the octant. 
The combination of these two channels provides a synergistic information on $\theta_{23}$.
A residual degeneracy persists between the $\theta_{23}$ octant and the CP-phase $\delta$, 
but it can be lifted with a balanced exposure of neutrino and antineutrino runs~\cite{Agarwalla:2013ju}.
 
Most of the octant sensitivity studies have been performed within the 3-flavor
framework (for recent works see~\cite{Agarwalla:2013ju,Agarwalla:2013hma,Bass:2013vcg,Bora:2014zwa,Das:2014fja,Nath:2015kjg}). 
However, there are strong indications that the standard picture may be incomplete.
In particular, there is a number of short-baseline (SBL) anomalies which point towards
the existence of new light eV-scale sterile neutrinos~\cite{Abazajian:2012ys,Palazzo:2013me,Gariazzo:2015rra}.
If this hypothesis gets confirmed, it may have 
important consequences for the 3-flavor searches. In fact, 
it has been already shown that the analysis of current LBL data is substantially affected 
by sterile neutrinos~\cite{Klop:2014ima,Palazzo:2015gja} and that the
discovery reach of CPV and MH of prospective LBL data can be substantially 
deteriorated~\cite{Agarwalla:2016mrc,Agarwalla:2016xxa} albeit, 
at least for the future experiment DUNE, a residual sensitivity at an 
acceptable ($\sim 3 \sigma$) level is preserved~\cite{Agarwalla:2016xxa}.
Recent studies of sterile neutrinos at LBL experiments can be found in~\cite{Hollander:2014iha,Berryman:2015nua,Gandhi:2015xza,Choubey:2016fpi}.

In this letter we point out that the situation is much worse for the $\theta_{23}$ octant, whose
determination is put in serious danger by the presence of active-sterile neutrino oscillations.
Taking DUNE as a case study, we show that in the 3+1 scheme, the newly identified 
interference term~\cite{Klop:2014ima} that appears in the $\nu_\mu \to \nu_e$ transition
probability can mimic a swap of the $\theta_{23}$ octant.  As a consequence, for unfavorable
combinations of the CP-phases in the 4-flavor scheme, the sensitivity to the octant 
of $\theta_{23}$ can be completely lost.

{\bf {\em Theoretical framework.}} In the 3+1 scheme,  a fourth mass eigenstate $\nu_4$ 
is introduced and the mixing is described by a $4\times4$ matrix 
\begin{equation}
\label{eq:U}
U =   \tilde R_{34}  R_{24} \tilde R_{14} R_{23} \tilde R_{13} R_{12}\,, 
\end{equation} 
where $R_{ij}$ ($\tilde R_{ij}$) is a real (complex) rotation in the ($i,j$) plane, 
which contains the $2\times2$ matrix 
\begin{eqnarray}
\label{eq:R_ij_2dim}
     R^{2\times2}_{ij} =
    \begin{pmatrix}
         c_{ij} &  s_{ij}  \\
         - s_{ij}  &  c_{ij}
    \end{pmatrix}
\,\,\,\,\,\,\,   
     \tilde R^{2\times2}_{ij} =
    \begin{pmatrix}
         c_{ij} &  \tilde s_{ij}  \\
         - \tilde s_{ij}^*  &  c_{ij}
    \end{pmatrix}
\,,    
\end{eqnarray}
in the $(i,j)$ sub-block. For brevity, we have introduced the definitions
\begin{eqnarray}
 c_{ij} \equiv \cos \theta_{ij} \qquad s_{ij} \equiv \sin \theta_{ij}\qquad  \tilde s_{ij} \equiv s_{ij} e^{-i\delta_{ij}}.
\end{eqnarray}

Let us now come to the transition probability relevant for the LBL experiment DUNE.
Matter effects have a sizable impact in DUNE, and confer a high sensitivity to the MH.
However, for simplicity, we neglect them in the considerations below, 
because they are basically irrelevant for the physical process that we want to highlight. We stress 
that in the numerical simulations we properly include the matter effects assuming a
line-averaged constant density of $2.87$\,g/cm$^3$ based on the PREM profile of Earth crust. 
In~\cite{Klop:2014ima}, it has been shown that the 4-flavor probability can be 
approximated by the sum of three terms
\begin{eqnarray}
\label{eq:Pme_4nu_3_terms}
P^{4\nu}_{\mu e}  \simeq  P_{\rm{0}} + P_{\rm {1}}+   P_{\rm {2}}\,,
\end{eqnarray}
which in vacuum take the form
\begin{eqnarray}
\label{eq:Pme_atm}
 & P_{\rm {0}} &\!\! \simeq\,  4 s_{23}^2 s^2_{13}  \sin^2{\Delta}\,,\\
\label{eq:Pme_int_1}
 & P_{\rm {1}} &\!\!  \simeq\,   8 s_{13} s_{12} c_{12} s_{23} c_{23} (\alpha \Delta)\sin \Delta \cos({\Delta \pm \delta_{13}})\,,\\
 \label{eq:Pme_int_2}
 & P_{\rm {2}} &\!\!  \simeq\,   4 s_{14} s_{24} s_{13} s_{23} \sin\Delta \sin (\Delta \pm \delta_{13} \mp \delta_{14})\,,
\end{eqnarray}
where $\Delta \equiv  \Delta m^2_{31}L/4E$ is the atmospheric oscillating frequency
depending on the baseline $L$ and the neutrino energy $E$, and
$\alpha \equiv \Delta m^2_{21}/ \Delta m^2_{31}$.
The double sign in front of the CP-phases
reflects the fact that it is opposite for neutrinos (upper sign) and antineutrinos
(lower sign). The first term $P_{\rm {0}}$, which is positive definite and 
independent of the CP-phases, gives the leading contribution to the probability.
The term $P_{\rm {1}}$ is related to the interference of the oscillations driven by the
solar and atmospheric frequencies. This term, apart from higher order corrections,
coincides with the standard interference term, which renders the 3-flavor transition probability 
sensitive to the CP-phase $\delta \equiv  \delta_{13}$.  The term  $P_{\rm {2}}$ is a genuine 
4-flavor effect, and is driven by the interference between the atmospheric frequency and the
 large frequency related to the new mass eigenstate~\cite{Klop:2014ima}.  This term does not manifest an 
 explicit dependency on $\Delta m^2_{41} \equiv m_4^2-m_1^2$ because the related 
 oscillations are very fast and get averaged out by the finite energy resolution of the detector.  
 We now observe that, as can be inferred from the latest 3-flavor global 
 analyses~\cite{Capozzi:2016rtj,Gonzalez-Garcia:2015qrr,Forero:2014bxa} and from the global 
 3+1 fits~\cite{Giunti:2013aea, Kopp:2013vaa}, the three small mixing angles have similar size 
 $s_{13} \sim s_{14} \sim s_{24} \simeq 0.15$, and therefore they can all be assumed of the same 
 order $\epsilon$, while the ratio  $\alpha \simeq \pm\, 0.03$ is of order $\epsilon^2$. This implies that
\begin{eqnarray}
 P_{\rm {0}} \sim \epsilon^2\,, \qquad
 P_{\rm {1}} \sim \epsilon^3\,, \qquad  
 P_{\rm {2}} \sim \epsilon^3 \,.
\end{eqnarray}
Now let us come to the $\theta_{23}$ octant issue. 
In general, we can re-express the atmospheric mixing angle as 
\begin{eqnarray}
 \label{eq:theta_23}
\theta_{23}  = \frac{\pi}{4} \pm \eta\,.
\end{eqnarray}
where $\eta$ is a positive-definite angle and the positive (negative) sign
corresponds to HO (LO). The current 3-flavor global fits~\cite{Capozzi:2016rtj,Gonzalez-Garcia:2015qrr,Forero:2014bxa} 
suggest that $\theta_{23}$ can deviate by no more than $\sim 6^0$ from maximal mixing. Equivalently,
 $\sin^2\theta_{23}$ must lie in the range $\sim [0.4,0.6]$. This implies that  $\eta$ is confined to relatively 
small values ($\eta \lesssim 0.1$), and can be considered of the same order of magnitude
($\epsilon$) of $s_{13}$, $s_{14}$ and $s_{24}$. Therefore, it is legitimate to use the expansion 
\begin{eqnarray}
 \label{eq:s2_theta_23}
s^2_{23} =  \frac{1}{2} (1 \pm \sin 2\eta\,)   \simeq \frac{1}{2} \pm \eta\,.
\end{eqnarray}
An experiment can be sensitive to the octant if, despite the freedom introduced by the unknown 
CP-phases, there is still a difference between the probabilities in the two octants, i.e.
\begin{eqnarray}
\label{eq:DPme}
\Delta P \equiv P^{\mathrm {HO}}_{\mu e} (\delta_{13}^{\mathrm {HO}}, \delta_{14}^{\mathrm {HO}}) -
                           P^{\mathrm {LO}}_{\mu e} (\delta_{13}^{\mathrm {LO}}, \delta_{14}^{\mathrm {LO}})\ne 0\,.
\end{eqnarray}
In Eq.~(\ref{eq:DPme}) we must think to one of the two octants as the true choice
(used to simulate data) and the other one as the test parameter (used to simulate the theoretical model).
For example, if for definiteness we fix the LO as the true choice, then
for a given combination of $(\delta_{13}^{\mathrm {LO}}, \delta_{14}^{\mathrm {LO}}$)
there can be sensitivity to the octant if  $\Delta P \ne 0$ in the hypothesis that
$(\delta_{13}^{\mathrm {HO}}, \delta_{14}^{\mathrm {HO}})$ are both unknown 
and free to vary in the range $[-\pi, \pi]$. According to Eq.~(\ref{eq:Pme_4nu_3_terms}), 
we can split  $\Delta P$ in the sum of three terms 
\begin{eqnarray}
\label{eq:DPme_4nu_3_terms}
\Delta P =   \Delta P_{\rm{0}} + \Delta P_{\rm {1}} +   \Delta P_{\rm {2}}\,.
\end{eqnarray}
The first term is positive-definite, does not depend on the CP-phases  and is given by 
\begin{eqnarray}
\label{eq:DP_0}
\Delta P_{\rm {0}} \simeq 8 \eta s_{13}^2 \sin^2\Delta\,.
\end{eqnarray}
The second and third terms depend on the CP-phases and can have both positive
or negative values. Their expressions are given by
\begin{eqnarray}
\label{eq:DP_1}
\Delta P_{\rm {1}}& =& A \big[ \cos(\Delta \pm \phi^{\mathrm {HO}}) - \cos(\Delta \pm \phi^{\mathrm {LO}})\big] \,,\\
\label{eq:DP_2}
\Delta P_{\rm {2}} &= &B \big[ \sin(\Delta \pm \psi^{\mathrm {HO}}) - \sin(\Delta \pm \psi^{\mathrm {LO}})\big]\,,
\end{eqnarray}
where for compactness, we have introduced the amplitudes%
\footnote{In the expressions of $A$ and $B$ we are neglecting terms proportional to powers 
of $\eta$, which would give rise to negligible (at least fourth order in $\epsilon$) corrections.}

\begin{eqnarray}
\label{eq:A}
A & =&  4 s_{13} s_{12} c_{12} (\alpha \Delta)\sin \Delta\,,\\
\label{eq:B}
B &= &  2 \sqrt{2} s_{14} s_{24} s_{13} \sin\Delta\,,
\end{eqnarray}
and the auxiliary CP-phases
\begin{eqnarray}
\label{eq:A}
\phi & =&  \delta_{13}\,,\\
\label{eq:B}
\psi &= &  \delta_{13} -  \delta_{14}\,,
\end{eqnarray}
with the appropriate superscripts (LO or HO). If we adopt as  a benchmark value
$\sin^2 \theta_{23} = 0.42\, (0.58)$ for LO (HO), i.e. $\eta = 0.08$, at the first oscillation
maximum ($\Delta = \pi/2$), we have 
\begin{eqnarray}
\label{eq:DP0_1}
\Delta P_{\rm {0}} \simeq   0.014\,, \quad|A| \simeq  0.013\,, \quad |B| \simeq  0.010\,.
\end{eqnarray}
These numbers give a feeling of the (similar) size of the three terms involved 
in Eq.~(\ref{eq:DPme_4nu_3_terms}). It is clear that an experiment can 
be sensitive to the octant only if the positive-definite difference $\Delta P_{\rm {0}}$ cannot be completely
compensated by a negative contribution coming from the sum of  $\Delta P_{\rm {1}}$ and $\Delta P_{\rm {2}}$.

\begin{figure}[t!]
\vspace*{-0.0cm}
\hspace*{-0.3cm}
\includegraphics[width=7.0 cm]{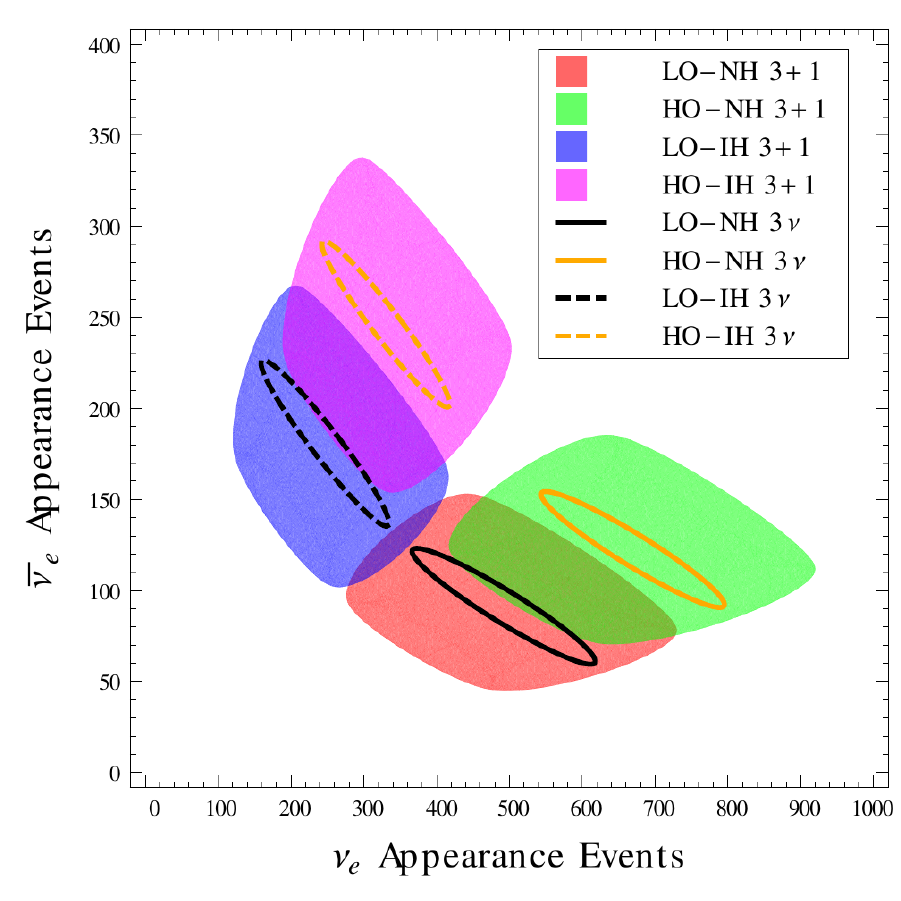}
\vspace*{-0.0cm}
\caption{Bi-event plot corresponding to the DUNE setup. The ellipses represent
the 3-flavor case, while the colored blobs correspond to the 3+1 scheme (see the legend). 
We take $\sin^2\theta_{23} = 0.42$\,(0.58) as  benchmark value for the LO (HO).
In the 3-flavor ellipses, the running parameter is $\delta_{13}$ varying in the range $[-\pi,\pi]$. 
In the 4-flavor blobs there are two running parameters, $\delta_{13}$ and $\delta_{14}$, both
varying in their allowed ranges $[-\pi,\pi]$. In the 3+1 case we have assumed $\theta_{34}$ = $0^0$.
\label{fig:bievents}}
\end{figure}

Now, we recall what happens in the 3-flavor framework, when
the last term $\Delta P_{\rm {2}}$ in Eq.~(\ref{eq:DPme_4nu_3_terms}) is absent. 
In this case, if one considers only the neutrino channel
there are unfavorable combinations of the two CP-phases $\delta_{13}^{\mathrm {LO}}$ and
 $\delta_{13}^{\mathrm {HO}}$ for which $\Delta P = 0$ and there is no sensitivity to the octant.
On the other hand, as recently recognized in~\cite{Agarwalla:2013ju,Machado:2013kya}, 
the octant-$\delta_{13}$ degeneracy can be lifted if one exploits also the antineutrino channel.
This fact can be understood from Fig.~\ref{fig:bievents}, which represents 
the bi-event plot for the DUNE experiment. In such a plot, the ellipses refer
to the 3-flavor case, while the colored blobs represent the 3+1 scheme.
We take $\sin^2 \theta_{23} = 0.42\,(0.58)$ as a benchmark value for
the LO (HO) octant. First, we notice that in DUNE the MH is not an issue because there
is a clear separation between NH and IH, which is basically guaranteed by the presence of matter effects%
\footnote{The small overlap between NH and IH blobs in Fig.~\ref{fig:bievents} for the LO case 
can be removed using the spectral information available in DUNE (see~\cite{Agarwalla:2016xxa}).}.
So we can fix the attention on one of the two hierarchies, for example the NH.
We observe, that the (black) ellipse corresponding to the LO is well separated
from the (yellow) HO ellipse. This separation is a synergistic effect of the fact
that we are considering {\em both} neutrino and antineutrino events.

Finally, let us come to the 3+1 scheme. In this case the third term in Eq.~(\ref{eq:DPme_4nu_3_terms}) 
is active. It depends on the additional CP-phase $\delta_{14}$, 
so its sign can been chosen independently of that of the second term. This circumstance
gives more freedom in the 3+1 scheme and there is more space for degeneracy.
The bi-event plot in Fig.~\ref{fig:bievents} confirms such a basic expectation. The graph
now becomes a blob, which can be seen as a convolution of an ensemble of ellipses 
(see~\cite{Agarwalla:2016mrc,Agarwalla:2016xxa}), and the separation between 
LO and HO is lost even if one considers both neutrino and antineutrino events. 
 
\begin{figure}[t!]
\vspace*{-0.5cm}
\hspace*{-0.5cm}
\includegraphics[width=9.5 cm]{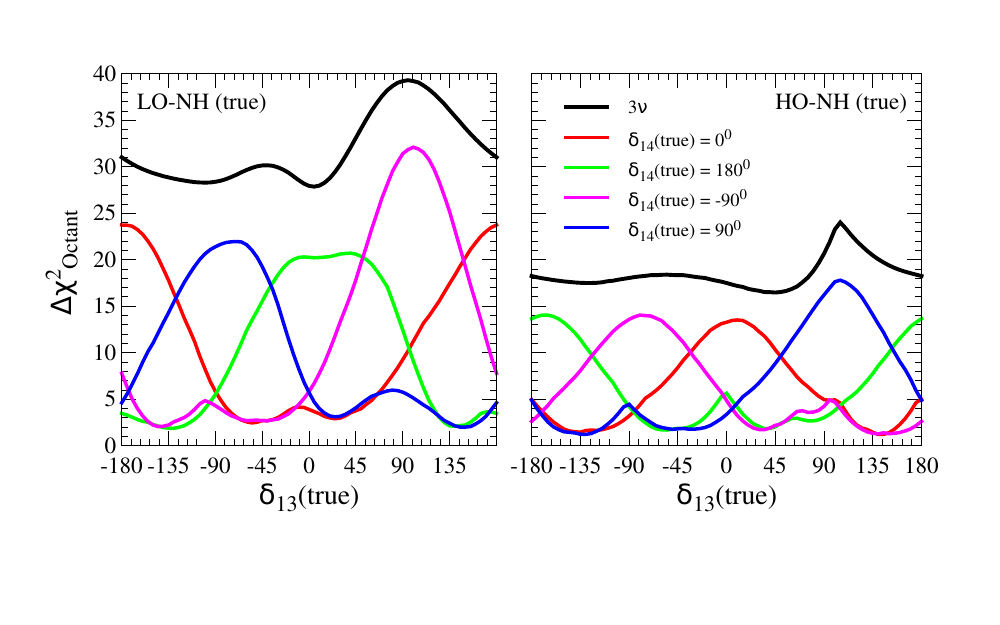}
\vspace*{-1.3cm}
\caption{Discovery potential for excluding the wrong octant as a function of true $\delta_{13}$ 
assuming LO-NH (left panel) and HO-NH (right panel) as the true choice. 
We take $\sin^2\theta_{23} = 0.42$\,(0.58) as  benchmark value for the LO (HO). 
In each panel, we present the results for the 3-flavor case (black line), and for the 3+1 scheme considering
four different values of true $\delta_{14}$ (colored lines). In the 3$\nu$ case, we marginalize
away ($\theta_{23}, \delta_{13}$) (test). In the 3+1 scheme, we fix $\theta_{14} = \theta_{24} = 9^0$, $\theta_{34} = 0$, 
and marginalize over ($\theta_{23}, \delta_{13}, \delta_{14}$) (test). 
\label{fig:2pan_octant_sens}}
\end{figure}

{\bf {\em  Numerical results.}} In our simulations, we use the GLoBES software~\cite{Huber:2004ka,Huber:2007ji}.
For DUNE, we consider a total exposure of 248 kt $\cdot$ MW $\cdot$ year, shared equally between neutrino  
and antineutrino modes. For the details of the DUNE setup, and of the statistical analysis, 
we refer the reader to our recent paper~\cite{Agarwalla:2016xxa}, and references therein. 
Figure~\ref{fig:2pan_octant_sens} displays  the discovery potential for identifying the true octant 
as a function of true $\delta_{13}$. The left (right) panel refers to the true choice LO-NH (HO-NH). 
In both panels, for comparison, we show the results for the 3-flavor case (represented by the black curve). 
Concerning the 3+1 scheme, we draw the curves corresponding to four representative values of 
true $\delta_{14}$ ($0^0, 180^0, -90^0, 90^0$). In the 3$\nu$ case we marginalize over ($\theta_{23}, \delta_{13}$) (test). 
In the 3+1 scheme, we fix $\theta_{14} = \theta_{24} = 9^0$ and $\theta_{34} = 0$, and 
we marginalize over ($\theta_{23}, \delta_{13}, \delta_{14}$) (test). In all cases we marginalize over the MH. 
However, we have checked that the minimum of $\Delta \chi^2$ is never reached in the wrong hierarchy. 
This confirms that  MH is not a source of degeneracy in the determination of the octant.

The 3-flavor curves have  been already discussed in the literature 
(see for example~\cite{Agarwalla:2013ju,Agarwalla:2013hma,Nath:2015kjg}). 
Nonetheless, we deem it useful to make the following remarks: 
i) a good $\theta_{23}$ octant sensitivity for all values of $\delta_{13}$ (true) can be
achieved with equal neutrino and antineutrino runs~\cite{Agarwalla:2013ju}, 
ii) the spectral information plays an important role in distinguishing between the two octants
for unfavorable choices of true hierarchy and $\delta_{13}$, and
iii) always the sensitivity is higher for LO  compared to HO irrespective 
of the hierarchy choice.  For the first time, during this work, we realized 
that this last issue of asymmetric sensitivity between LO and HO is related
to a synergistic effect of the $\nu_\mu \to \nu_\mu$ and $\nu_\mu \to \nu_e$ channels.
Basically, the $\nu_\mu  \to \nu_\mu$ channel fixes the test value of $\theta_{23}$ in the 
octant opposite to its true value. However, such a test value is not exactly equal to its  
octant symmetric choice (i.e., $\theta_{23}^{\rm {test}} \ne \pi/2 - \theta_{23}^{\rm{true}}$).
This happens because the $\nu_\mu \to \nu_\mu$ survival probability contains higher order 
octant-sensitive terms, which can been probed in high-statistics experiments like DUNE. 
We find that these corrections always go in the direction of increasing
(decreasing) the difference $|\sin^2 \theta_{23}^{\mathrm{test}} - \sin^2\theta_{23}^{\mathrm{true}}|$
by $\sim 15 \%$ with respect to its default value $|0.58-0.42| = 0.16$ in the LO (HO) case.
Since the leading term of the $\nu_\mu \to \nu_e$ appearance channel is 
sensitive to this difference, the performance is enhanced (suppressed) in the LO (HO) case.

Fig.~\ref{fig:2pan_octant_sens} shows that in the 3+1 scheme there exist unfavorable 
combinations of $\delta_{13}$ (true) and $\delta_{14}$ (true) for which the octant sensitivity
falls below the 2$\sigma$ level. We have verified that for such combinations the spectra
corresponding to the two octants are almost indistinguishable both for neutrinos and 
antineutrinos. Therefore, even a broad-band experiment such as DUNE 
cannot break the degeneracy introduced by a sterile neutrino. 

\begin{figure}[t!]
\vspace*{-0.5cm}
\hspace*{-0.5cm}
\includegraphics[width=9.5 cm]{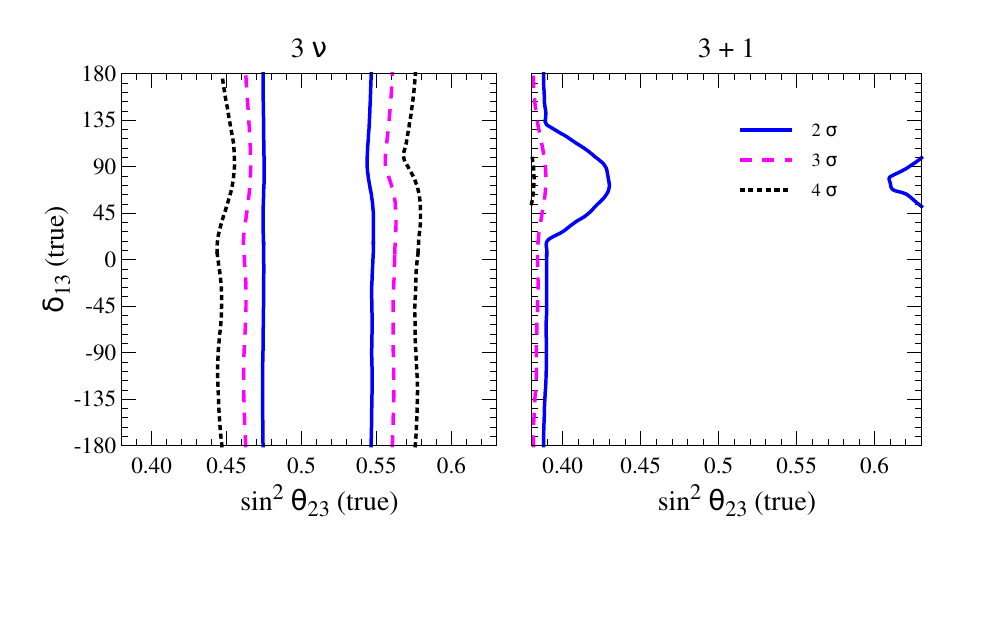}
\vspace*{-1.3cm}
\caption{Discovery potential for excluding the wrong octant in [$\sin^2 \theta_{23}, \delta_{13}$] (true) plane
assuming NH as true choice.
The left (right) panel corresponds to the 3$\nu$ (3+1) case.
In the 3-flavor case, we marginalize away ($\theta_{23}, \delta_{13}$) (test). In the 3+1 case, 
in addition, we marginalize over $\delta_{14}$ (true) and $\delta_{14}$ (test) fixing
$\theta_{14} = \theta_{24} = 9^0$ and $\theta_{34} = 0$. The solid blue, 
dashed magenta, and dotted black curves correspond, respectively, to the 
2$\sigma$, 3$\sigma$, and 4$\sigma$ confidence levels (1 d.o.f.).
\label{fig:2pan_octant_true}}
\end{figure}

So far, we have considered two true values
of $\sin^2\theta_{23} = 0.42$ (LO) and  0.58 (HO) (see Fig.~\ref{fig:2pan_octant_sens}). 
However, it is interesting to ask how things change if different choices are made for the
true value of $\theta_{23}$. Figure~\ref{fig:2pan_octant_true} answers this question.
It represents  the discovery potential for identifying the true octant in the plane
[$\sin^2 \theta_{23}, \delta_{13}$] (true) assuming NH as true choice.
The left (right) panel corresponds to the 3$\nu$ (3+1) scheme. 
In the 3+1 case we marginalize  away also the CP-phase $\delta_{14}$ (true) 
(in addition to all the test parameters) since it is unknown. Hence, the outcome of this
procedure determines the minimal  guaranteed sensitivity, i.e. the one corresponding 
to the worst case scenario. The solid blue, dashed magenta, and dotted black curves correspond, 
respectively, to 2$\sigma$, 3$\sigma$ and 4$\sigma$ confidence levels (1 d.o.f.). The comparison 
of the two panels gives a bird-eye view of the situation. It is clear that, in the 3+1 scheme, no minimal 
sensitivity is guaranteed in the entire plane. We have checked that similar conclusions are valid also
in the case of IH as true MH.

{\bf {\em Conclusions.}} In this letter, we have addressed for the first time 
the impact of a light eV-scale sterile neutrino in identifying the octant of the 
mixing angle $\theta_{23}$ at the next generation LBL experiment DUNE. 
We have shown that in the 3+1 scheme, the new recently identified interference 
term~\cite{Klop:2014ima} that enters the $\nu_\mu \to \nu_e$ transition probability 
can perfectly mimic a swap of the $\theta_{23}$ octant, even when we use the 
information from both the neutrino and antineutrino channels. As a consequence, 
the sensitivity to the octant of $\theta_{23}$ can be completely lost in the presence 
of active-sterile oscillations. It remains to be seen if other kinds of experiments, 
in particular those using atmospheric neutrinos, can lift or at least alleviate the degeneracy 
that we have found in the context of LBL experiments. Our educated guess is that this would 
prove to be very difficult since obtaining a satisfying $\theta_{23}$ octant sensitivity 
with atmospheric neutrinos is a very hard task already in the 3-flavor framework, 
and in the enlarged 3+1 scheme, the situation should naturally worsen. 
At the end, we just want to emphasize the fact that the presence of light sterile neutrinos 
will have far-reaching consequences on the physical effects that we are going to observe 
in future long-baseline experiments such as DUNE, and we may land up with a different 
interpretation of the measured event spectra.

\section*{Acknowledgments}

S.K.A. is supported by the DST/INSPIRE Research Grant [IFA-PH-12],
Department of Science \& Technology, India. A part of S.K.A.'s work was
carried out at the International Centre for Theoretical Physics (ICTP), 
Trieste, Italy. It is a pleasure for him to thank the ICTP for the hospitality
and support during his visit via SIMONS Associateship. A.P. is supported 
by the grant ``Future In Research''  {\it Beyond three neutrino families}, 
Fondo di Sviluppo e Coesione 2007-2013, 
APQ Ricerca Regione Puglia, Italy, ÒProgramma regionale a sostegno della
specializzazione intelligente e della sostenibilit\`a sociale ed ambientale. 
A.P. acknowledges partial support by the research project 
{\em TAsP} funded by the Instituto Nazionale di Fisica Nucleare (INFN). 

\section*{Note added}

After the completion of this work, the IceCube~\cite{TheIceCube:2016oqi}, 
Daya Bay~\cite{An:2016luf}, and MINOS~\cite{MINOS:2016viw} Collaborations
reported new constraints on active-sterile mixing. The new upper limit from the 
IceCube Collaboration~\cite{TheIceCube:2016oqi} on $\sin^22\theta_{24}$ around the current best-fit 
$\Delta m^2_{41}$ $\simeq$ 1.75 eV$^2$~\cite{Collin:2016aqd} is $\simeq$ 0.2
at 99\% confidence level, which means that $\theta_{24}$ 
is constrained to be smaller than $13^{\circ}$ or so. The MINOS Collaboration 
placed a new upper limit of $\simeq 0.03$ on $\sin^2\theta_{24}$ around 
$\Delta m^2_{41}$ $\simeq$ 1.75 eV$^2$ at 90\% C.L.~\cite{MINOS:2016viw,Adamson:2016jku}
suggesting that the upper limit on $\theta_{24}$ is around $10^{\circ}$. 
Therefore, the benchmark value $\theta_{24}$ = $9^{\circ}$ which we have considered 
in our work is compatible with these new limits. The combined analysis of the Daya Bay and Bugey-3
data provides a new constraint~\cite{Adamson:2016jku}
on $\sin^22\theta_{14}$ which is $\simeq$ 0.06 at 90\% confidence level 
around $\Delta m^2_{41} \simeq 1.75\,\, {\mathrm {eV}}^2$. 
This implies that $\theta_{14}$ is smaller than $7^{\circ}$ or so, which
is pretty close to the benchmark value of $\theta_{14} = 9^{\circ}$ that we have considered in our analysis.
For completeness, we have explicitly checked performing new simulations that our results remain 
almost unaltered if we take $\theta_{14}$ = $7^{\circ}$ and $\theta_{24}$ = $9^{\circ}$
instead of our benchmark choice of $\theta_{14} = \theta_{24} = 9^{\circ}$.
Hence, we can safely conclude that
the sensitivity towards the octant of $\theta_{23}$ can be completely lost even in light of
the new constraints on the active-sterile mixing that were reported recently.

\bibliographystyle{h-physrev4}
\bibliography{Sterile-References}

\begin{thebibliography}{10}

\bibitem{Pascoli:2013wca}
S.~Pascoli and T.~Schwetz,
\newblock Adv.High Energy Phys. {\bf 2013}, 503401 (2013).

\bibitem{Agarwalla:2013hma}
S.~K. Agarwalla, S.~Prakash and S.~Uma~Sankar,
\newblock JHEP {\bf 1403}, 087 (2014), [1304.3251].

\bibitem{Agarwalla:2014fva}
S.~K. Agarwalla,
\newblock Adv.High Energy Phys. {\bf 2014}, 457803 (2014), [1401.4705].

\bibitem{Feldman:2012qt}
G.~Feldman, J.~Hartnell and T.~Kobayashi,
\newblock Adv.High Energy Phys. {\bf 2013}, 475749 (2013), [1210.1778].

\bibitem{Stanco:2015ejj}
L.~Stanco,
\newblock 1511.09409.

\bibitem{Mohapatra:2006gs}
R.~Mohapatra and A.~Smirnov,
\newblock Ann.Rev.Nucl.Part.Sci. {\bf 56}, 569 (2006), [hep-ph/0603118].

\bibitem{Albright:2006cw}
C.~H. Albright and M.-C. Chen,
\newblock Phys.Rev. {\bf D74}, 113006 (2006), [hep-ph/0608137].

\bibitem{Altarelli:2010gt}
G.~Altarelli and F.~Feruglio,
\newblock Rev. Mod. Phys. {\bf 82}, 2701 (2010), [1002.0211].

\bibitem{King:2014nza}
S.~F. King, A.~Merle, S.~Morisi, Y.~Shimizu and M.~Tanimoto,
\newblock New J. Phys. {\bf 16}, 045018 (2014), [1402.4271].

\bibitem{King:2015aea}
S.~F. King,
\newblock J. Phys. {\bf G42}, 123001 (2015), [1510.02091].

\bibitem{Merle:2014eja}
A.~Merle, S.~Morisi and W.~Winter,
\newblock JHEP {\bf 07}, 039 (2014), [1402.6332].

\bibitem{Rivera-Agudelo:2015vza}
D.~C. Rivera-Agudelo and A.~P{\'e}rez-Lorenzana,
\newblock Phys. Rev. {\bf D92}, 073009 (2015), [1507.07030].

\bibitem{Fukuyama:1997ky}
T.~Fukuyama and H.~Nishiura,
\newblock hep-ph/9702253.

\bibitem{Mohapatra:1998ka}
R.~N. Mohapatra and S.~Nussinov,
\newblock Phys.Rev. {\bf D60}, 013002 (1999), [hep-ph/9809415].

\bibitem{Lam:2001fb}
C.~Lam,
\newblock Phys.Lett. {\bf B507}, 214 (2001), [hep-ph/0104116].

\bibitem{Harrison:2002et}
P.~Harrison and W.~Scott,
\newblock Phys.Lett. {\bf B547}, 219 (2002), [hep-ph/0210197].

\bibitem{Kitabayashi:2002jd}
T.~Kitabayashi and M.~Yasue,
\newblock Phys.Rev. {\bf D67}, 015006 (2003), [hep-ph/0209294].

\bibitem{Grimus:2003kq}
W.~Grimus and L.~Lavoura,
\newblock Phys.Lett. {\bf B572}, 189 (2003), [hep-ph/0305046].

\bibitem{Koide:2003rx}
Y.~Koide,
\newblock Phys.Rev. {\bf D69}, 093001 (2004), [hep-ph/0312207].

\bibitem{Mohapatra:2005yu}
R.~Mohapatra and W.~Rodejohann,
\newblock Phys.Rev. {\bf D72}, 053001 (2005), [hep-ph/0507312].

\bibitem{Ma:2002ge}
E.~Ma,
\newblock Mod.Phys.Lett. {\bf A17}, 2361 (2002), [hep-ph/0211393].

\bibitem{Ma:2001dn}
E.~Ma and G.~Rajasekaran,
\newblock Phys.Rev. {\bf D64}, 113012 (2001), [hep-ph/0106291].

\bibitem{Babu:2002dz}
K.~Babu, E.~Ma and J.~Valle,
\newblock Phys.Lett. {\bf B552}, 207 (2003), [hep-ph/0206292].

\bibitem{Grimus:2005mu}
W.~Grimus and L.~Lavoura,
\newblock JHEP {\bf 0508}, 013 (2005), [hep-ph/0504153].

\bibitem{Ma:2005mw}
E.~Ma,
\newblock Mod.Phys.Lett. {\bf A20}, 2601 (2005), [hep-ph/0508099].

\bibitem{Raidal:2004iw}
M.~Raidal,
\newblock Phys.Rev.Lett. {\bf 93}, 161801 (2004), [hep-ph/0404046].

\bibitem{Minakata:2004xt}
H.~Minakata and A.~Y. Smirnov,
\newblock Phys.Rev. {\bf D70}, 073009 (2004), [hep-ph/0405088].

\bibitem{Ferrandis:2004vp}
J.~Ferrandis and S.~Pakvasa,
\newblock Phys.Rev. {\bf D71}, 033004 (2005), [hep-ph/0412038].

\bibitem{Antusch:2005ca}
S.~Antusch, S.~F. King and R.~N. Mohapatra,
\newblock Phys.Lett. {\bf B618}, 150 (2005), [hep-ph/0504007].

\bibitem{Hall:1999sn}
L.~J. Hall, H.~Murayama and N.~Weiner,
\newblock Phys.Rev.Lett. {\bf 84}, 2572 (2000), [hep-ph/9911341].

\bibitem{deGouvea:2012ac}
A.~de~Gouvea and H.~Murayama,
\newblock 1204.1249.

\bibitem{Fogli:1996pv}
G.~L. Fogli and E.~Lisi,
\newblock Phys. Rev. {\bf D54}, 3667 (1996), [hep-ph/9604415].

\bibitem{Barger:2001yr}
V.~Barger, D.~Marfatia and K.~Whisnant,
\newblock Phys. Rev. {\bf D65}, 073023 (2002), [hep-ph/0112119].

\bibitem{Minakata:2001qm}
H.~Minakata and H.~Nunokawa,
\newblock JHEP {\bf 10}, 001 (2001), [hep-ph/0108085].

\bibitem{Hiraide:2006vh}
K.~Hiraide {\em et~al.},
\newblock Phys.Rev. {\bf D73}, 093008 (2006), [hep-ph/0601258].

\bibitem{BurguetCastell:2002qx}
J.~Burguet-Castell, M.~Gavela, J.~Gomez-Cadenas, P.~Hernandez and O.~Mena,
\newblock Nucl.Phys. {\bf B646}, 301 (2002), [hep-ph/0207080].

\bibitem{Agarwalla:2013ju}
S.~K. Agarwalla, S.~Prakash and S.~U. Sankar,
\newblock JHEP {\bf 1307}, 131 (2013), [1301.2574].

\bibitem{Machado:2013kya}
P.~Machado, H.~Minakata, H.~Nunokawa and R.~Z. Funchal,
\newblock 1307.3248.

\bibitem{Minakata:2013hgk}
H.~Minakata and S.~J. Parke,
\newblock Phys. Rev. {\bf D87}, 113005 (2013), [1303.6178].

\bibitem{Chatterjee:2013qus}
A.~Chatterjee, P.~Ghoshal, S.~Goswami and S.~K. Raut,
\newblock JHEP {\bf 1306}, 010 (2013), [1302.1370].

\bibitem{Bass:2013vcg}
M.~Bass {\em et~al.},
\newblock Phys. Rev. {\bf D91}, 052015 (2015), [1311.0212].

\bibitem{Bora:2014zwa}
K.~Bora, D.~Dutta and P.~Ghoshal,
\newblock Mod. Phys. Lett. {\bf A30}, 1550066 (2015), [1405.7482].

\bibitem{Das:2014fja}
C.~R. Das, J.~Maalampi, J.~Pulido and S.~Vihonen,
\newblock JHEP {\bf 02}, 048 (2015), [1411.2829].

\bibitem{Nath:2015kjg}
N.~Nath, M.~Ghosh and S.~Goswami,
\newblock 1511.07496.

\bibitem{Abazajian:2012ys}
K.~N. Abazajian {\em et~al.},
\newblock 1204.5379.

\bibitem{Palazzo:2013me}
A.~Palazzo,
\newblock Mod. Phys. Lett. {\bf A28}, 1330004 (2013), [1302.1102].

\bibitem{Gariazzo:2015rra}
S.~Gariazzo, C.~Giunti, M.~Laveder, Y.~F. Li and E.~M. Zavanin,
\newblock 1507.08204.

\bibitem{Klop:2014ima}
N.~Klop and A.~Palazzo,
\newblock Phys. Rev. {\bf D91}, 073017 (2015), [1412.7524].

\bibitem{Palazzo:2015gja}
A.~Palazzo,
\newblock Phys. Lett. {\bf B757}, 142 (2016), [1509.03148].

\bibitem{Agarwalla:2016mrc}
S.~K. Agarwalla, S.~S. Chatterjee, A.~Dasgupta and A.~Palazzo,
\newblock JHEP {\bf 02}, 111 (2016), [1601.05995].

\bibitem{Agarwalla:2016xxa}
S.~K. Agarwalla, S.~S. Chatterjee and A.~Palazzo,
\newblock 1603.03759.

\bibitem{Hollander:2014iha}
D.~Hollander and I.~Mocioiu,
\newblock Phys. Rev. {\bf D91}, 013002 (2015), [1408.1749].

\bibitem{Berryman:2015nua}
J.~M. Berryman, A.~de~Gouv{\^e}a, K.~J. Kelly and A.~Kobach,
\newblock Phys. Rev. {\bf D92}, 073012 (2015), [1507.03986].

\bibitem{Gandhi:2015xza}
R.~Gandhi, B.~Kayser, M.~Masud and S.~Prakash,
\newblock JHEP {\bf 11}, 039 (2015), [1508.06275].

\bibitem{Choubey:2016fpi}
S.~Choubey and D.~Pramanik,
\newblock 1604.04731.

\bibitem{Capozzi:2016rtj}
F.~Capozzi, E.~Lisi, A.~Marrone, D.~Montanino and A.~Palazzo,
\newblock 1601.07777.

\bibitem{Gonzalez-Garcia:2015qrr}
M.~C. Gonzalez-Garcia, M.~Maltoni and T.~Schwetz,
\newblock 1512.06856.

\bibitem{Forero:2014bxa}
D.~V. Forero, M.~Tortola and J.~W.~F. Valle,
\newblock Phys. Rev. {\bf D90}, 093006 (2014), [1405.7540].

\bibitem{Giunti:2013aea}
C.~Giunti, M.~Laveder, Y.~F. Li and H.~W. Long,
\newblock Phys. Rev. {\bf D88}, 073008 (2013), [1308.5288].

\bibitem{Kopp:2013vaa}
J.~Kopp, P.~A.~N. Machado, M.~Maltoni and T.~Schwetz,
\newblock JHEP {\bf 05}, 050 (2013), [1303.3011].

\bibitem{Huber:2004ka}
P.~Huber, M.~Lindner and W.~Winter,
\newblock Comput.Phys.Commun. {\bf 167}, 195 (2005), [hep-ph/0407333].

\bibitem{Huber:2007ji}
P.~Huber, J.~Kopp, M.~Lindner, M.~Rolinec and W.~Winter,
\newblock Comput.Phys.Commun. {\bf 177}, 432 (2007), [hep-ph/0701187].

\bibitem{TheIceCube:2016oqi}
IceCube, M.~G. Aartsen {\em et~al.},
\newblock Phys. Rev. Lett. {\bf 117}, 071801 (2016), [1605.01990].

\bibitem{An:2016luf}
Daya Bay, F.~P. An {\em et~al.},
\newblock Phys. Rev. Lett. {\bf 117}, 151802 (2016), [1607.01174].

\bibitem{MINOS:2016viw}
MINOS, P.~Adamson {\em et~al.},
\newblock Phys. Rev. Lett. {\bf 117}, 151803 (2016), [1607.01176].

\bibitem{Collin:2016aqd}
G.~H. Collin, C.~A. Arg{\"u}elles, J.~M. Conrad and M.~H. Shaevitz,
\newblock Phys. Rev. Lett. {\bf 117}, 221801 (2016), [1607.00011].

\bibitem{Adamson:2016jku}
MINOS, Daya Bay, P.~Adamson {\em et~al.},
\newblock Phys. Rev. Lett. {\bf 117}, 151801 (2016), [1607.01177],
\newblock [Addendum: Phys. Rev. Lett.117,no.20,209901(2016)].

\end{thebibliography}

\end{document}